\documentclass[10pt, conference]{IEEEtran}
\usepackage{pifont}
\usepackage{array}
\usepackage{booktabs}
\usepackage{multirow}
\usepackage{subfigure}
\usepackage{times,amsmath,color,amssymb,amsfonts,graphicx,epsfig,cite,psfrag,subfigure,algorithm,epstopdf}
\usepackage{algorithm}
\usepackage{algpseudocode}
\usepackage{textcomp}
\def\BibTeX{{\rm B\kern-.05em{\sc i\kern-.025em b}\kern-.08em
    T\kern-.1667em\lower.7ex\hbox{E}\kern-.125emX}}

\newcommand\eps\varepsilon

\begin{document}


\def\FIGDIR{./figs}

\IEEEoverridecommandlockouts

\title{Collaborative Knowledge Sharing-empowered Effective Semantic Rate Maximization for Two-tier Semantic-Bit Communication Networks}

\author{\IEEEauthorblockN{Hong Chen, Fang Fang, and Xianbin Wang\\}
\IEEEauthorblockA{Department of Electrical and Computer Engineering,
Western University, London, ON, CANADA\\ E-mail: {\{hche88, fang.fang, xianbin.wang\}@uwo.ca}}}

%
%

\maketitle \thispagestyle{empty}

\begin{abstract}

Effective task-oriented semantic communications relies on perfect knowledge alignment between transmitters and receivers for accurate recovery of task-related semantic information, which can be susceptible to knowledge misalignment and performance degradation in practice. To tackle this issue, continual knowledge updating and sharing are crucial to adapt to evolving task and user related demands, despite the incurred resource overhead and increased latency.
In this paper, we propose a novel collaborative knowledge sharing-empowered semantic transmission mechanism in a two-tier edge network, exploiting edge cooperations and bit communications to address KB mismatch. By deriving a generalized effective semantic transmission rate (GESTR) that considers both semantic accuracy and overhead, we formulate a mixed integer nonlinear programming problem to maximize GESTR of all mobile devices by optimizing knowledge sharing decisions, extraction ratios, and BS/subchannel allocations, subject to task accuracy and delay requirements. The joint optimum solution can be obtained by proposed fractional programming based branch and bound algorithm and modified Kuhn-Munkres algorithm efficiently. Simulation results demonstrate the superior performance of proposed solution, especially in low signal-to-noise conditions.

\end{abstract}


\sloppy
\allowdisplaybreaks

\vspace{-1mm}
\section{Introduction}
\label{sec:introduction}

As an intelligent communication paradigm, task-oriented semantic communications aim to exchange task-related semantic information, rather than the raw information bits from the communication source, which reveals great potential to enhance the communication efficiency and alleviate the spectrum scarcity \cite{task1,task2}. Achieving effective semantic communication requires perfect alignment of task-related knowledge within the knowledge bases (KBs) of both transmitters and receivers, however, which is not always feasible in practice \cite{Hu2023Robust}. Mismatched knowledge can impede the accurate reconstruction of semantic information, thereby degrading the overall performance of semantic communication systems.

A few works \cite{Sun2024Semantic,MS} proposed different transmission schemes to consider the KB mismatch issue between transmitters and receivers.
Different levels of features can be transmitted based on KB matching conditions in a KB-aware multi-level feature transmission framework \cite{Sun2024Semantic} for remote zero-shot object recognition. Authors in \cite{MS} developed a queuing model in semantic communications, considering both knowledge-match and knowledge-mismatch packets to derive the packet loss probability.
However, these works did not consider the knowledge updating and sharing between transmitters and receivers, which is crucial to maintain the perfect KB alignment and facilitate adaptation to new tasks and user demands \cite{GAI}.

Knowledge updating and sharing also compete for limited radio resources that could impact semantic transmission given the strict task delay demands. Especially when channel condition is bad, knowledge updating and sharing from mobile devices (MDs) with limited power cannot always be guaranteed. In this case, the collaboration of edge servers with powerful capabilities becomes critically significant. Bit communications can also be used as a supplemental method to transmit the remaining data related to unshared knowledge, ensuring the effective task execution, which is considered in our paper.

Semantic extraction ratio has been considered in resource-constrained networks, integrating computing offloading \cite{Zheng2023Computing}, power and bandwidth allocations \cite{Cang2023Online,Liu2024Adaptable}, and user selection \cite{Liu2024Adaptable}. These works aim to optimize traditional network performance, such as task delay \cite{Zheng2023Computing}, energy consumption \cite{Cang2023Online}, and task success probability \cite{Liu2024Adaptable}. Considering the semantic overhead introduced by knowledge sharing transmission latency, a semantic-level performance metric is derived and studied instead in our paper. Additionally, these works assume that the MD has the perfect KB alignment with the edge server, which is not always feasible due to network dynamics.

In this paper, we propose a novel collaborative knowledge sharing-empowered semantic transmission mechanism between heterogeneous transceivers in a two-tier edge network, leveraging edge cooperations and bit communications to tackle the challenge of KB mismatch. A generalized effective semantic transmission rate (GESTR) is derived considering semantic accuracy and overhead. The total GESTR of all MDs is maximized by jointly optimizing knowledge sharing decisions, extraction ratios, and BS/subchannel allocations, considering semantic accuracy and delay requirements of tasks. The formulated mixed integer nonlinear programming (MINLP) problem is decomposed equivalently and solved by proposed fractional programming based branch and bound (FP-BnB) algorithm and modified Kuhn-Munkres (K-M) algorithm optimally and efficiently. A variety of results demonstrate the superior performance of proposed mechanism and optimum solution, especially when signal-to-noise ratio of MDs is low.


\vspace{-2.5mm}
\section{System Model and Problem Formulation}
\label{sec:systemmodel}

\begin{figure}[t]
  \centering
  \includegraphics[width=68mm]{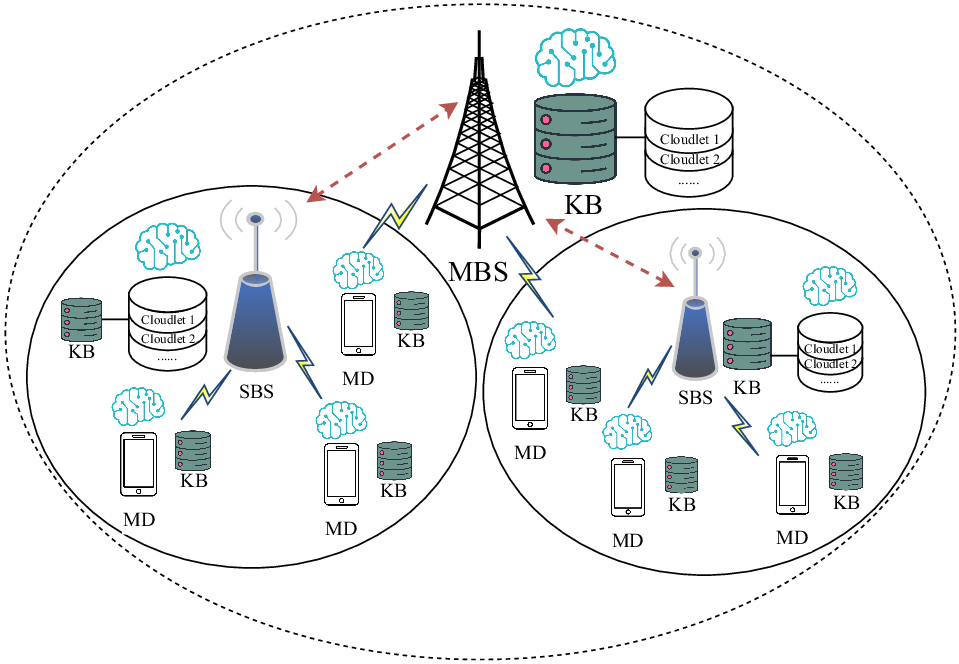} \vspace{-3mm}
  \caption{Two-tier edge network architecture for KB-aware task-oriented semantic-bit communications.}
  \label{fig:1}
\vspace{-3mm}
\end{figure}

As shown in Fig. \ref{fig:1}, we consider a two-tier edge network, where the first tier consists of one macro base station (MBS) co-located with a powerful edge server and the second tier comprises multiple small base stations (SBSs), each of witch is equipped with an edge server and coexists within the coverage of MBS. The MBS can communicate with the SBSs via wireless backhaul links.
Both base stations (BSs) and MDs can perform semantic communications based on the shared knowledge. Each MD is equipped with a pre-trained semantic encoder to transmit the task-related semantic information to the associated BS. A KB is formed at each MD to store common and private knowledge. While each BS is equipped with a pre-trained semantic decoder to recovery the received semantic information, and several cloudlets at edge server to execute the target tasks of MDs. A KB is also established at each BS and needs to be updated according to the target tasks in the current network.
We consider $I$ MDs in the coverage of SBSs, which indicates the MDs are also in the coverage of MBS. MD is indexed by $i \in {\cal I}=\{1,2,\ldots,I\}$. BS is indexed by $j \in {\cal J}=\{0,1,\ldots,J\}$, where $j=0$ represents the MBS and $j=1,2,\ldots,J$ indicates the SBSs. There have $K$ available subchannels, which is indexed by $k \in {\cal K}=\{1,2,\ldots,K\}$.
Each MD can access MBS or SBS to transmit the requested data to complete its target task at the edge server of either BS. Define $\delta_{i,j,k}\in \{0,1\}$ indicating if MD $i$ is associated with BS $j$ using subchannel $k$. If $\delta_{i,j,k}=1$, MD $i$ is associated with BS $j$ on subchannel $k$; otherwise, $\delta_{i,j,k}=0$.
Define ${\cal L}_i=\{1,2,\ldots,L_i\} $ as the class set of task-related knowledge in $L_i$ classes of MD $i$, which is indexed by $l \in {\cal L}_i$. As BSs have their own KBs, some classes of MD task-related knowledge may be already stored, e.g., some common knowledge. The initial set of task-related knowledge classes of MD $i$ stored at BS $j$ is denoted by ${\cal L}_{i,j}^{\rm int} = \{1,2,\ldots,{L}_{i,j}^{\rm int}\}$.
If ${\cal L}_{i,j}^{\rm int} = {\cal L}_i$, i.e., task-related knowledge at both transmitter and receiver is well-matched, MD $i$ can start semantic communications immediately with BS $j$. However, if ${\cal L}_{i,j}^{\rm int} \subsetneq {\cal L}_i$, there exists mismatched task-related knowledge classes of MD $i$ at the KB of BS $j$, denoted by ${\cal L}_{i,j}^{\rm mis} = {\cal L}_i \setminus {\cal L}_{i,j}^{\rm int}$.

\begin{figure}[t]
\centering
  \includegraphics[width=68mm]{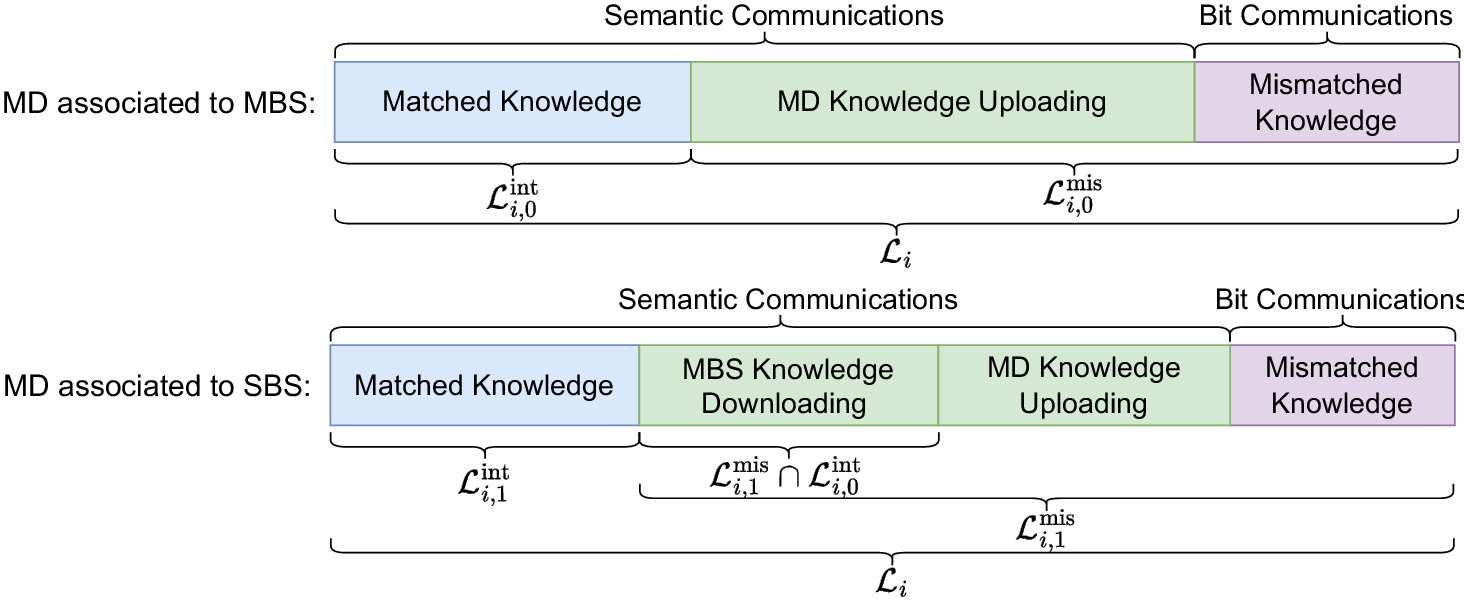} \vspace{-3.5mm}
  \caption{An illustration of proposed collaborative knowledge sharing-empowered semantic-bit transmission mechanism when ${\cal L}_{i,1}^{\rm int} \subsetneq {\cal L}_{i}$.}
  \label{fig:2}
\vspace{-6.5mm}
\end{figure}

\vspace{-2mm}
\subsection{Collaborative knowledge sharing-empowered transmission}

In order to overcome the KB mismatch issue, we propose a novel collaborative knowledge sharing-empowered semantic-bit transmission mechanism, as shown in Fig. \ref{fig:2}. When a MD is associated with MBS, the mismatched task-related knowledge classes of MD can be shared and uploaded to associated MBS. While a MD is associated with an SBS, the mismatched knowledge classes can be uploaded by the MD and downloaded from MBS via wireless backhaul link, if MBS has the mismatched knowledge classes stored, i.e., $l \in {\cal L}_{i,j}^{\rm mis} \cap {\cal L}_{i,0}^{\rm int}, \forall j \neq 0$. This collaborative method can improve the semantic performance when channel condition between the MD and SBS is bad and the MD is in low-battery status. Nevertheless, due to task delay tolerance and random channel conditions, all the mismatched knowledge may not be able to be shared to the KB of associated BS. The requested source data related to unshared knowledge classes has to be transmitted in traditional bit communications as a form of compensation. We define $a_{i,j,k,l}\in \{0,1\}$ as the transmission mode indicator. If $a_{i,j,k,l}=1$, the requested data related to knowledge class $l$ of MD $i$ associated with BS $j$ is transmitted in semantic communications on subchannel $k$. Otherwise, $a_{i,j,k,l}=0$, the requested data is transmitted in bit communications. In order to realize semantic communications, there exist two knowledge sharing manners for classes $ l \in {\cal L}_{i,j}^{\rm mis} \cap {\cal L}_{i,0}^{\rm int}$ of a MD associated with SBS. We let $b_{i,j,k,l} \in \{0,1\}, \forall j \neq 0$ be the knowledge sharing manner indicator. If $b_{i,j,k,l}=1$, mismatched knowledge class $l$ of MD $i$ is uploaded from MD $i$ to SBS $j$ on subchannel $k$; otherwise, if $b_{i,j,k,l}=0$, mismatched class $l$ of MD $i$ is downloaded from the MBS to SBS $j$ on subchannel $k$. The knowledge classes $ l \in {\cal L}_{i,j}^{\rm mis} \setminus {\cal L}_{i,0}^{\rm int}$ can only be uploaded by the MD, i.e., $b_{i,j,k,l}=1$. Note that when a MD is associated with the MBS, $b_{i,0,k,l}=1, \forall l$.
We define $d_{i,l}^{\rm T}$ and $d_{i,l}^{\rm K}$ as the size of requested source data and related knowledge data at MD $i$ in knowledge class $l$, respectively. Let $I_{i,l}$ and $c_{i,l}$ be the amount of semantic information and computing load of requested data related to knowledge class $l$ of MD $i$, correspondingly.

\vspace{-2.8mm}
\subsection{Target task completion time}

If MD $i$ is associated with SBS $j$, the mismatched knowledge classes can be uploaded from the MD or downloaded from the MBS.The knowledge uploading time can be formulated as
\vspace{-3mm}
\begin{equation}\label{UP}
  t_{i,j,k}^{\rm up} \!= \!\frac{\sum\limits_{l \in {\cal L}_{i,j}^{\rm mis}\cap {\cal L}_{i,0}^{\rm int}} \!\! a_{i,j,k,l}b_{i,j,k,l}{d_{i,l}^{\rm K}}+\!\!\!\! \sum\limits_{l \in {\cal L}_{i,j}^{\rm mis} \setminus {\cal L}_{i,0}^{\rm int}} \!\! a_{i,j,k,l}{d_{i,l}^{\rm K}}}{R_{i,j,k}},
  \vspace{-2mm}
\end{equation}
for $j \neq 0$, where the physical transmission still follows the Shannon theory and $R_{i,j,k}$ is the bit transmission rate from MD $i$ to BS $j$ on subchannel $k$, given by
\vspace{-2mm}
\begin{equation}\label{rate}
  R_{i,j,k} = W{\log_2}( 1 + \frac{{p^{\rm{T}}_i}{g_{i,j,k}}}{\sigma^2}),
  \vspace{-2mm}
\end{equation}
where $W$ denotes the subchannel bandwidth, $p^{\rm{T}}_i$ and $g_{i,j,k}$ are the transmit power and the link gain from MD $i$ to BS $j$ using subchannel $k$, correspondingly, and $\sigma^2$ is the noise power at the BS receiver input.

While the knowledge downloading time is formulated as
\vspace{-2mm}
\begin{equation}\label{DOWN}
  t_{i,j,k}^{\rm down} = \frac{\sum\limits_{l \in {\cal L}_{i,j}^{\rm mis}\cap {\cal L}_{i,0}^{\rm int}} a_{i,j,k,l}(1-b_{i,j,k,l}){d_{i,l}^{\rm K}}}{R_{i,k}^{0 \rightarrow j}},
  \vspace{-2mm}
\end{equation}
for $j \neq 0$, where $R_{i,k}^{0 \rightarrow j}$ is the MBS transmission rate to SBS $j$ connecting with MD $i$ on subchannel $k$, which can be calculated by
\vspace{-4mm}
\begin{equation}\label{rateMBS}
  R_{i,k}^{0 \rightarrow j }= W{\log_2}( 1 + \frac{{p^{\rm{T}}_{0,j}}{g_{i,k}^{0 \rightarrow j}}}{\sigma^2}),
  \vspace{-2mm}
\end{equation}
for $j \neq 0$, where ${p^{\rm{T}}_{0,j}}$ is the transmit power of MBS to SBS $j$ and ${g_{i,k}^{0 \rightarrow j}}$ represents the link gain from the MBS to SBS $j$ connecting with MD $i$ on subchannel $k$. To prevent incurring interference degrading the transmission performance, each MD associated with only one BS can access one and only one available subchannel, and each subchannel can be occupied by one and only one MD, i.e.,
\vspace{-2mm}
\begin{equation}\label{BA}
  \sum\limits_{j \in {\cal J}}\sum\limits_{k \in {\cal K}} \delta_{i,j,k} \le 1,
  \vspace{-2mm}
\end{equation}
\begin{equation}\label{SA}
  \sum\limits_{i \in {\cal I}}\sum\limits_{j \in {\cal J}} \delta_{i,j,k} \le 1.
  \vspace{-2mm}
\end{equation}
Besides, when the mismatched knowledge classes of MD $i$ at SBS is downloaded from the MBS, the MBS shares the same subchannel occupied by MD $i$ via adopting time division multiple access, i.e., the MBS and MD $i$ use subchannel $k$ successively.

If MD $i$ is associated with the MBS, the mismatched knowledge classes at MBS can be uploaded from the MD. The knowledge uploading time can be formulated as
\vspace{-2mm}
\begin{equation}\label{UPMBS}
  t_{i,j,k}^{\rm up} = \frac{\sum\limits_{l \in {\cal L}_{i,j}^{\rm mis}} a_{i,j,k,l}{d_{i,l}^{\rm K}}}{R_{i,j,k}},
  \vspace{-2mm}
\end{equation}
for $j=0$. Note that the knowledge uploading time in this case is same as \eqref{UP} when $b_{i,j,k,l}=1, \forall l$.

Thus, the knowledge sharing time for target task of MD $i$ associated with BS $j$ on subchannel $k$ can be given by
\vspace{-2mm}
\begin{equation}\label{share}
  t_{i,j,k}^{\rm K} = \left\{
                        \begin{array}{ll}
                          t_{i,j,k}^{\rm up}+t_{i,j,k}^{\rm down}, & {j \neq 0;} \\
                          t_{i,j,k}^{\rm up}, & {j=0.}
                        \end{array}
                      \right.
\end{equation}

Let $\xi_{i,j,k}\in [0,1]$ be the semantic extraction ratio of requested source data from MD $i$ to BS $j$ on subchannel $k$.
The semantic transmission time can be expressed as
\vspace{-2mm}
\begin{equation}
  t_{i,j,k}^{\rm S} = \frac{\xi_{i,j,k} (\sum\limits_{l \in {\cal L}_{i,j}^{\rm int}} {d_{i,l}^{\rm T}}+ \sum\limits_{l \in {\cal L}_{i,j}^{\rm mis}} {a_{i,j,k,l} d_{i,l}^{\rm T}}  )}{R_{i,j,k}}.
  \vspace{-2mm}
\end{equation}
The requested data related to unshared mismatched knowledge classes can only be transmitted in bit communications for effective task execution. The bit transmission time can be formulated as
\vspace{-4.5mm}
\begin{equation}
  t_{i,j,k}^{\rm B} = \frac{\sum\limits_{l \in {\cal L}_{i,j}^{\rm mis}} (1-a_{i,j,k,l}) {d_{i,l}^{\rm T}}}{R_{i,j,k}}.
  \vspace{-2mm}
\end{equation}

To execute target task, BS will assign one of the cloudlets after receiving the requested data from the MD. Additional computing load is required to process the semantic data rather than source data, due to the extra computations for semantic decoding and reconstruction \cite{Cang2023Online}. We define $\omega_{i,j,k} \ge 1$ as the ratio of computing load of semantic data to that of source data. Without loss of generality, we assume that $
  \omega_{i,j,k}= \frac{1}{{\xi_{i,j,k}}^{\rho}}$,
where $\rho > 0$ is a constant parameter varying with adopted semantic models to different task types. Note that $\omega_{i,j,k} = 1$ when source data is processed, i.e., $\xi_{i,j,k} = 1$. Hence, the computing time of semantic data can be formulated as
\vspace{-2mm}
\begin{equation}\label{CT1}
  t_{i,j,k}^{\rm R} = \frac{\omega_{i,j,k} ( \sum\limits_{l \in {\cal L}_{i,j}^{\rm int}} {c_{i,l}}+ \sum\limits_{l \in {\cal L}_{i,j}^{\rm mis}} {a_{i,j,k,l} c_{i,l}})}{f_j^{\rm C}},
  \vspace{-2mm}
\end{equation}
where ${f_j^{\rm C}}$ is the computing speed of a cloudlet at BS $j$ in number of CPU cycles per second. While the computing time for executing source data can be given by
\vspace{-2mm}
\begin{equation}\label{CT2}
  t_{i,j,k}^{\rm E} = \frac{\sum\limits_{l \in {\cal L}_{i,j}^{\rm mis}} (1-a_{i,j,k,l}) c_{i,l}}{f_j^{\rm C}}.
  \vspace{-3mm}
\end{equation}

Thus, the task completion time can be obtained as
\vspace{-2mm}
\begin{equation}\label{TT}
  t_{i,j,k} = t_{i,j,k}^{\rm K} +t_{i,j,k}^{\rm S} + t_{i,j,k}^{\rm B} + t_{i,j,k}^{\rm R}+ t_{i,j,k}^{\rm E}.
  \vspace{-2mm}
\end{equation}
The target task of MD $i$ must be completed within maximum delay tolerance $t_i^{\max}$, i.e.,
\vspace{-2mm}
\begin{equation}\label{Delay}
  \sum\limits_{j \in {\cal J}}\sum\limits_{k \in {\cal K}} {\delta_{i,j,k} t_{i,j,k}} \le t_i^{\max}.
  \vspace{-2mm}
\end{equation}

\subsection{Generalized effective semantic transmission rate}
\label{sec:STR}

Semantic transmission accuracy $\eps_{i,j,k}$ depends on the semantic extraction ratio $\xi_{i,j,k}$. Since deriving a closed-form formula for $\eps_{i,j,k}(\xi_{i,j,k})$ is intractable due to the unexplainability of neural networks of semantic models, the semantic accuracy in general can be characterized as a nonlinear function of semantic extraction ratio \cite{Xu2024Task,Liu2024Adaptable}, i.e.,
\vspace{-2mm}
\begin{align}\label{accuracyfunction}
  \eps_{i,j,k}(\xi_{i,j,k}) &\approx \eps'_{i,j,k}(\xi_{i,j,k}|\theta_1,\theta_2,\theta_3,\theta_4) \nonumber \\
  &=-\theta_1  e^{(\theta_2 (1-\xi_{i,j,k}))}+\theta_3 e^{(-\theta_4 (1-\xi_{i,j,k}))},
\end{align}

\vspace{-3mm}\noindent
where $\theta_1,\theta_2,\theta_3,\theta_4 \ge 0$ are tuning parameters varying with adopted semantic models to different task types. The optimal parameters $(\theta_1,\theta_2,\theta_3,\theta_4)$ can be found via nonlinear least squares fitting \cite{gavin2019levenberg}. Since the semantic information of source data is transmitted with less data, the reconstructed information is less accurate. Thus, the semantic accuracy is a monotonically increasing function of semantic extraction ratio.

Each MD $m$ has a minimum semantic accuracy requirement $\eps_m^{\rm th}$ to guarantee the semantic transmission performance, i.e.,
\vspace{-2mm}
\begin{equation}\label{accuracy}
  \sum\limits_{j \in {\cal J}}\sum\limits_{k \in {\cal K}} \delta_{i,j,k} \eps_{i,j,k}(\xi_{i,j,k}) \ge \eps_i^{\rm th}.
  \vspace{-2mm}
\end{equation}

Unlike bit-stream data rate, the semantic unit (sut) as the basic unit of semantic information can be used to measure the amount of semantic information \cite{Yan2022Resource}.
Effective semantic transmission rate, as one of the crucial semantic-based performance metrics, is defined as the effectively transmitted semantic information per second in suts/s.
Moreover, knowledge sharing, essential for enabling effective semantic communications, is considered semantic overhead and factored into the effective semantic transmission rate, with considerations of semantic accuracy.
In our proposed mechanism, the requested data could be transmitted in two manners of semantic and bit communications.
Thus, we derive a GESTR $\gamma_{i,j,k}$ considering semantic overhead and accuracy as
\vspace{-2mm}
\begin{equation}\label{STR}
  \gamma_{i,j,k}= \frac{\eps_{i,j,k}(\!\!\sum\limits_{l \in {\cal L}_{i,j}^{\rm int}} \!\!{I_{i,l}}+ \!\!\!\! \sum\limits_{l \in {\cal L}_{i,j}^{\rm mis}} \!\!{a_{i,j,k,l} I_{i,l}}) + \!\!\!\! \sum\limits_{l \in {\cal L}_{i,j}^{\rm mis}} {\!\!\!\!(1\!-\!a_{i,j,k,l}) I_{i,l}}}{t_{i,j,k}^{\rm K} +t_{i,j,k}^{\rm S} + t_{i,j,k}^{\rm B}}.
  \vspace{-1.5mm}
\end{equation}

In order to reduce spectrum pressure and address KB mismatch challenge in a two-tier edge network, our goal is to maximize the total GESTR of all MDs by jointly optimizing knowledge sharing decisions, semantic extraction ratios, and BS/subchannel allocations, considering semantic overhead, semantic accuracy and delay requirements of MD target tasks. The optimization problem can be formulated as follows:
\vspace{-2mm}
\begin{subequations} \label{Eq:maxtotal}
\begin{align}
\max_{\delta_{i,j,k}, a_{i,j,k,l}, \atop b_{i,j,k,l}, \xi_{i,j,k}} & \sum\limits_{i \in {\cal I}}\sum\limits_{j \in {\cal J}}\sum\limits_{k \in {\cal K}} {\delta_{i,j,k}  \gamma_{i,j,k} }   \\
\text{s.t.} ~~~~~& \eqref{BA}, \eqref{SA}, \eqref{Delay}, \eqref{accuracy},   \label{Eq:C1} \\
& \delta_{i,j,k}, a_{i,j,k,l}, b_{i,j,k,l}  \in \{0,1\}, \forall i,j,k,l  \label{Eq:C2}\\
& \xi_{i,j,k} \in [0,1], \forall i,j,k. \label{Eq:C3}
\end{align}
\end{subequations}

\vspace{-2mm}\noindent
Problem \eqref{Eq:maxtotal} belongs to the class of MINLP problem, which is NP-hard and cannot be solved efficiently using traditional optimization methods. Next, we will develop a joint optimum algorithm to find the optimal solution to problem \eqref{Eq:maxtotal}.

\vspace{-2.5mm}
\section{Joint Optimum Solution to the Problem}
\label{sec:solution}

Due to the coupled and mixed variables and the fractional format of objective function, it is difficult to find the optimum solution using traditional tools. In this section, problem \eqref{Eq:maxtotal} is decomposed into multiple joint subproblems and an allocation subproblem. An optimum algorithm using FP-BnB algorithm is proposed to find the optimal solution efficiently to joint subproblems. The optimal solutions of subproblems are then fed into the allocation subproblem, which is solved by a modified K-M algorithm \cite{KM2} optimally.

\vspace{-3mm}
\subsection{Joint subproblems}
\label{subsec:jointsub}

Given $\delta_{i,j,k}=1$, i.e., MD $i$ is associated with BS $j$ on subchannel $k$, the reduced subproblem is independent and can be separated from problem \eqref{Eq:maxtotal} as follows:
\vspace{-2mm}
\begin{subequations} \label{Eq:maxone}
\begin{align}
\max_{a_{i,j,k,l}, b_{i,j,k,l}, \xi_{i,j,k}} & {\gamma_{i,j,k} }   \\
\text{s.t.} ~~~~~~~&  t_{i,j,k} \le t_i^{\max},   \label{Eq:CC1} \\
& \eps_{i,j,k}(\xi_{i,j,k}) \ge \eps_i^{\rm th}, \label{Eq:CC2} \\
& a_{i,j,k,l}, b_{i,j,k,l}  \in \{0,1\}, \forall l  \label{Eq:CC3}\\
& \xi_{i,j,k} \in [0,1]. \label{Eq:CC4}
\end{align}
\end{subequations}

\vspace{-2mm}\noindent
Subproblem \eqref{Eq:maxone} is still an MINLP problem. Fortunately, we observe that continuous variable $\xi_{i,j,k}$ is in a finite range $[0,1]$. Since $\eps_{i,j,k}(\xi_{i,j,k})$ is a monotonically increasing function, it is easy to obtain $\xi_{i,j,k} \ge \xi_{i,j,k}^{\rm th}$ from \eqref{accuracyfunction}, where $\xi_{i,j,k}^{\rm th}$ is the minimum semantic extraction ratio satisfying minimum semantic accuracy constraint \eqref{Eq:CC2}. Thus, one-dimension linear search of $\xi_{i,j,k}$ can be conducted in range $[\xi_{i,j,k}^{\rm th}, 1]$ via breaking it into $M$ equal segments, so that $\xi_{i,j,k}$ takes values $\xi_{i,j,k}^{(m)} = \xi_{i,j,k}^{(m-1)}+ (1-\xi_{i,j,k}^{\rm th})/M$, for $m=1,2,\ldots,M$ and $\xi_{i,j,k}^{(0)}=\xi_{i,j,k}^{\rm th}$.
With fixed $\xi_{i,j,k}^{(m)}$, the resulted problem can be rewritten as follows:
\vspace{-3mm}
\begin{subequations} \label{Eq:givenxi}
\begin{align}
\max_{{\bf a}, {\bf b}} & \frac{X({\bf a})}{Y({\bf a}, {\bf b})}   \\
\text{s.t.} ~&  T({\bf a}, {\bf b}) \le t_i^{\max},   \label{Eq:CCC1} \\
& {\bf a}, {\bf b} \in \{0,1\}^{L_{i,j}^{\rm mis}},   \label{Eq:CCC2}
\end{align}
\end{subequations}

\vspace{-2mm}\noindent
where ${\bf a}\!=\![a_{i,j,k,l}, \forall l]$ and ${\bf b}\!=\![b_{i,j,k,l}, \forall l]$, $X\!({\bf a})$ and $Y\!({\bf a}, {\bf b})$ are the functions in the numerator and denominator of objective function, respectively, and $T\!({\bf a}, {\bf b})$ is the task completion time function.
Note that we give the solution assuming MD $i$ is associated with an SBS, since when an MD is associated with the MBS, $b_{i,j,k,l}\!=\!1, \forall l$.
To solve problem \eqref{Eq:givenxi}, we need to address the non-linear property that arises from the coupled ${\bf a}{\bf b}$ and the fractional format of objective function. We will show that the relaxation of problem \eqref{Eq:givenxi} can be transformed into a linear programming by FP so that problem \eqref{Eq:givenxi} can be solved optimally by proposed FP-BnB algorithm.

BnB algorithm can find the optimal solution for integer programming if the relaxed problem is a linear/convex problem, by branching it into smaller subproblems and using bounds to eliminate subproblems that cannot contain the optimal solution.
By introducing an auxiliary variable ${\bf \tilde{b}} = [\tilde{b}_{i,j,k,l} = a_{i,j,k,l}b_{i,j,k,l}, \forall l ]$, problem \eqref{Eq:givenxi} can be rewritten equivalently:
\vspace{-2mm}
\begin{subequations} \label{Eq:tildeb}
\begin{align}
\max_{{\bf a}, {\bf \tilde{b}}} & \frac{X({\bf a})}{Y({\bf a}, {\bf \tilde{b}})}   \\
\text{s.t.} ~&  T({\bf a}, {\bf \tilde{b}}) \le t_i^{\max},   \label{Eq:CCCC1} \\
& 0 \le [{\bf \tilde{b}}]_l \le [{\bf a}]_l, \forall l \label{Eq:CCCC2} \\
& {\bf a}, {\bf \tilde{b}} \in \{0,1\}^{L_{i,j}^{\rm mis}},   \label{Eq:CCCC3}
\end{align}
\end{subequations}

\vspace{-2.5mm}\noindent
where $[\cdot]_l$ indicates the $l$-th element of a vector.

{\bf Relaxation and branching:} By relaxing all variables ${\bf a}, {\bf \tilde{b}} \in \{0,1\}^{L_{i,j}^{\rm mis}}$ into relaxed variables ${\bf a}', {\bf \tilde{b}}' \in [0,1]^{L_{i,j}^{\rm mis}}$, we apply BnB method on the relaxation of problem \eqref{Eq:tildeb} to branch it into subproblems by iteratively fixing one fractional solution into binary values.
Thus, branching along would amount to brute-force enumeration of candidate solutions but with keeping track of bounds and pruning the search space, the algorithm can eliminate candidate solutions that it can prove will not contain an optimal solution.

{\bf Fractional programming:} At each branch $n$, the obtained relaxed subproblem with partial integer solutions and relaxed variables is still non-linear non-convex but can be addressed by FP to find the optimal solution. Denote the remaining relaxed variables at branch $n$ as ${\bf a'}^{(n)}, {\bf \tilde{b'}}^{(n)}$.
The relaxed subproblem can be transformed as
\vspace{-3mm}
\begin{subequations} \label{Eq:FP}
\begin{align}
\max_{{\bf a'}^{(n)}, {\bf \tilde{b'}}^{(n)}} & {X({\bf a'}^{(n)})}-\eta^{(n)}{Y({\bf a'}^{(n)}, {\bf \tilde{b'}}^{(n)})}   \\
\text{s.t.} ~~~&  T({\bf a'}^{(n)}, {\bf \tilde{b'}}^{(n)}) \le t_i^{\max},   \label{Eq:CCCCC1} \\
& 0 \le [{\bf \tilde{b'}}^{(n)}]_l \le [{\bf a'}^{(n)}]_l, \forall l \label{Eq:CCCCC2} \\
& {\bf a'}^{(n)}, {\bf \tilde{b'}}^{(n)} \in [0,1]^{L_{i,j}^{\rm mis (n)}},   \label{Eq:CCCCC3}
\end{align}
\end{subequations}

\vspace{-2.5mm}\noindent
with a new auxiliary variable $\eta^{(n)}$, iteratively updated by
\vspace{-2.5mm}
\begin{equation}\label{eta}
  \eta^{(n)}[q+1]=\frac{X({\bf a'}^{(n)}[q])}{Y({\bf a'}^{(n)}[q], {\bf \tilde{b'}}^{(n)}[q])},
  \vspace{-2mm}
\end{equation}
where $q$ is the iteration index.
When $\eta^{(n)}$ is fixed, \eqref{Eq:FP} becomes a linear programming and can be solved conveniently. After \eqref{Eq:FP} is solved, $\eta^{(n)}$ is recalculated, and the problem is solved again based on updated $\eta^{(n)}$. This process is repeated until it converges, when the optimum solutions are obtained (line 6), i.e., the optimum solutions to relaxation of \eqref{Eq:tildeb} at branch $n$. The iterative process is given in Algorithm \ref{algo1}, where $o$ is a small value as the tolerable error for convergence, and a proof of convergence is presented in \cite{Shen2018}.

{\bf Pruning and bounding:} The depth-first search is adopted, where the lower bound is defined as the maximum objective value of the current feasible integer solutions.
At each branch, the optimum objective value (if feasible) can be obtained and if smaller than the lower bound, this branch is discarded, same as the case if infeasible. Else, continue to the next branch.
The BnB algorithm can systematically reduce the search space by strategically pruning the branches according to the updated bounds, leading to a global optimum solution efficiently.
After obtaining optimum solution ${\bf a}^*, {\bf \tilde{b}}^*$, optimum solution ${\bf b}^*$ can be obtained based on the definition of ${\bf \tilde{b}}$.

\vspace{-3mm}
\subsection{Allocation subproblem}
\label{subsec:allosub}

After obtaining optimum solutions $a_{i,j,k,l}^*$'s, $b_{i,j,k,l}^*$'s and $\xi_{i,j,k}^*$'s for all $i,j,k$, we feed them into the original problem \eqref{Eq:maxtotal} and obtain the following allocation subproblem,
\vspace{-2.5mm}
\begin{subequations} \label{Eq:maxdelta}
\begin{align}
\max_{\delta_{i,j,k}} & \sum\limits_{i \in {\cal I}}\sum\limits_{j \in {\cal J}}\sum\limits_{k \in {\cal K}} {\delta_{i,j,k}  \gamma_{i,j,k}^* }   \\
\text{s.t.} ~& \eqref{BA}, \eqref{SA},    \label{Eq:1} \\
& \delta_{i,j,k}  \in \{0,1\}, \forall i,j,k  \label{Eq:2}
\end{align}
\end{subequations}

\vspace{-2.5mm}\noindent
which is a binary linear programming and can be regarded as an optimum matching problem in a bipartite graph. It can be solved by modified K-M algorithm optimally, where two vertex sets are ${\cal I}$ and ${\cal K}$ respectively, and $\arg\max_{j\in {\cal J}} \gamma_{i,j,k}^*$ is set as the weight between MD $i$ and subchannel $k$.

\begin{algorithm}[t]
\caption{Iterative algorithm for FP at branch $n$} \label{algo1}
\small{
\begin{algorithmic}[1]
\State $\eta^{(n)}=0$
\Repeat
    \State Solve \eqref{Eq:FP} with fixed $\eta^{(n)}$ to obtain the optimum ${\bf a'}^{(n)*}, {\bf \tilde{b'}}^{(n)*}$
\If {${{X({\bf a'}^{(n)*})}-\eta^{(n)}{Y({\bf a'}^{(n)*}, {\bf \tilde{b'}}^{(n)*})}} \le o$}
	\State Convergence = {\bf true}
    \State\Return $\eta^{(n)*}=\frac{X({\bf a'}^{(n)*})}{Y({\bf a'}^{(n)*}, {\bf \tilde{b'}}^{(n)*})}$
\Else
    \State $\eta^{(n)}=\frac{X({\bf a'}^{(n)*})}{Y({\bf a'}^{(n)*}, {\bf \tilde{b'}}^{(n)*})}$
\EndIf
\Until {Convergence = {\bf true}}
\end{algorithmic}
}
\end{algorithm}

\vspace{-2.5mm}
\section{Simulation Results}
\label{sec:simulation}

In this section, simulation results are presented to demonstrate the superior performance of proposed joint optimum algorithm compared with no collaboration scheme \cite{Xu2024Task} and another no knowledge sharing scheme \cite{KBC}.
We consider an SBS having a circular service area with a radius of 150 m centered at the origin, which is under the coverage of an MBS located at (-150 m, 0). The network has 5 subchannels and 3 MDs, and the locations of all MDs are uniformly distributed in the service area. There are 10 knowledge classes in the system, where the KBs at MBS and SBS store 6 and 5 classes randomly picked from 10 classes, respectively, and each MD requires 6 knowledge classes.
The link gains consider both distance-based path loss and small-scale fading, given as $g_{i,j,k}=10^{-3}{\rho_{i,j,k}}^2 d_{i,j}^{-2}$, where $d_{i,j}$ is the distance between MD $i$ and BS $j$ and ${\rho_{i,j,k}}^2$ is a random variable with exponential distribution and unit mean, since $\rho_{i,j,k}$ is the additional Rayleigh distributed small-scale fading \cite{Chen2024Semantic}.
The semantic accuracy nonlinear model \cite{Liu2024Adaptable} with $(\theta_1,\theta_2,\theta_3,\theta_4)=(-6.205e-8,16.45,0.9228,-0.06917)$ is adopted to estimate the relationship between semantic accuracy and extraction ratio.
Default parameters are summarized in Table~\ref{parameters}, where the values refer to \cite{Liu2024Adaptable,Chen2024Semantic} and $U[A,B]$ represents the uniform distribution between $A$ and $B$. The simulation results are obtained by averaging over 100 independent experiments, each of which is based on one set of randomly generated MD locations, and task and knowledge parameters.

\begin{figure}[t]
  \centering
  \includegraphics[height=50mm,width=60mm]{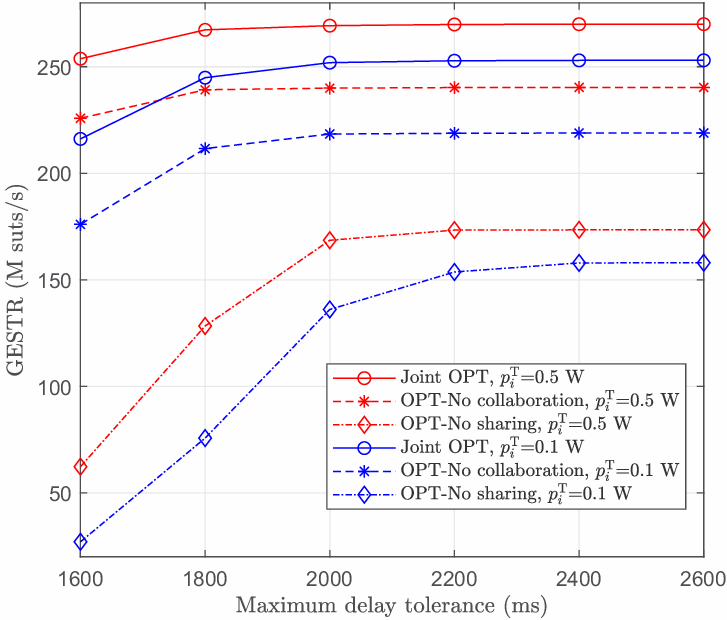} \vspace{-3mm}
  \caption{Total GESTR versus maximum delay tolerance.}
  \label{fig:3}
  \vspace{-7mm}
\end{figure}

Fig.~\ref{fig:3} shows the total GESTR of MDs versus the maximum delay tolerance $t_i^{\max}$ (same for all MDs). With the increase of $t_i^{\max}$, the total GESTR of all solutions increases significantly when $t_i^{\max}$ is relatively small. The increase becomes saturated when $t_i^{\max}$ is sufficiently large. It is because more mismatched knowledge classes can be shared to the KB of associated BS as the increase of $t_i^{\max}$, leading to higher GESTR; as $t_i^{\max}$ is sufficiently large, task delay tolerance can always be satisfied so that the performance is no longer affected by it. It further shows that the total GESTR of joint optimum solution is higher than that of optimum solution without collaboration and much higher than that of optimum solution without knowledge sharing, which verifies the effectiveness of proposed transmission mechanism and joint optimum solution. These results help demonstrate how much the joint collaborative knowledge sharing in proposed algorithm contributes to the superior performance of task-oriented hybrid communications. We can see that the total GESTR of all MDs when $p_i^{\rm T}=0.5$ W is higher than that when $p_i^{\rm T}=0.1$ W, since higher MD transmit power can reduce wireless transmission time. Moreover, the gap between proposed solution and the comparisons when $p_i^{\rm T}=0.1$ W is larger than that when $p_i^{\rm T}=0.5$ W, which further demonstrates the excellent performance of proposed mechanism and solution especially when the signal-to-noise ratio is relatively low.

\begin{table}[!t]
\begin{center}
\caption{Default Parameter Settings}
\label{parameters}
\scriptsize{
\begin{tabular}{|c|c|c|c|}
\hline
Parameters     &     Values  & Parameters     &     Values\\
\hline
$I_{i,l}$ &  $U[2,20]$ M suts/s & $f_n^{\rm C}$   &  [4, 2] G Hz\\
\hline
$d_{i,l}^{\rm K}$ &  $U[5,50]$ M bits & $p_i^{\rm{T}}$ &  0.1 W\\
\hline
$d_{i,l}^{\rm T}$ &  $U[20,100]$ M bits & ${p^{\rm{T}}_{0,j}}$ & 20 W \\
\hline
$c_{i,l}$ &  $U[1,100]$ M CPU cycles & $W$ & 6 M Hz\\
\hline
$\eps^{\rm th}_{i}$ &  $U[0.7,0.85]$ & $\sigma^2$ &  -120 dBm \\
\hline
$t^{\max}_{i}$ &  $U[2500,3500]$ ms & $\rho$   &  1  \\
\hline
\end{tabular}
}
\end{center}
\vspace{-2mm}
\end{table}

\begin{figure}[t]
  \centering
  \includegraphics[height=50mm,width=60mm]{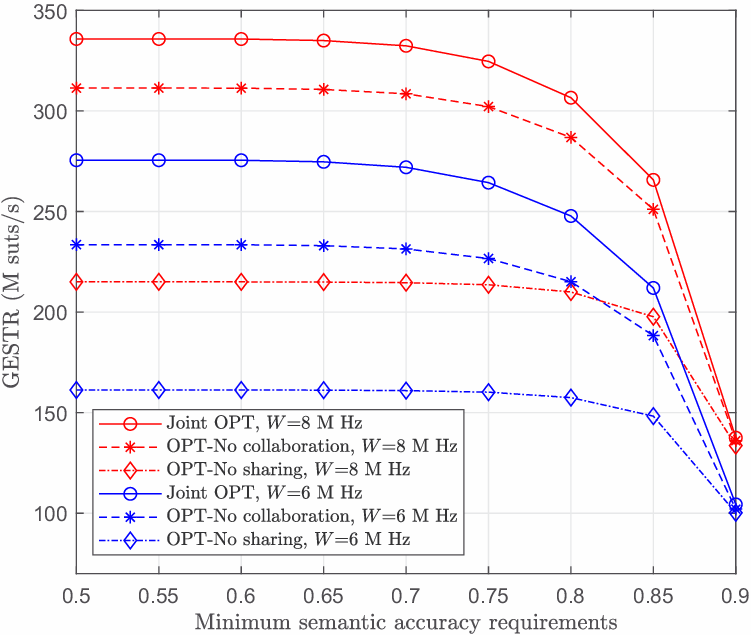} \vspace{-3mm}
  \caption{Total GESTR versus minimum semantic accuracy requirements.}
  \label{fig:4}
  \vspace{-7.7mm}
\end{figure}

Fig.~\ref{fig:4} shows the total GESTR of all MDs versus the minimum semantic accuracy requirements $\eps_i^{\rm th}$ (same for all MDs). The total GESTR is a constant when $\eps_i^{\rm th}$ is relatively small and then drops gradually as $\eps_i^{\rm th}$ increases. More semantic data should be extracted from source data to meet higher semantic accuracy requirement, resulting in longer semantic transmission time.
It also can be seen that the proposed solution has superior performance compared to the optimum solutions without considering edge collaboration and knowledge sharing. The total GESTR of all MDs for all solutions when $W=8$ M Hz is higher than that when $W=6$ M Hz. However, the gap between proposed solution and no collaboration scheme when $W=6$ M Hz is larger than that when $W=8$ M Hz, which reveals the advantages of proposed collaborative knowledge sharing mechanism when wireless channel condition is relatively bad.

\vspace{-2mm}
\section{Conclusions}
\label{sec:conclusions}

In this paper, we proposed a collaborative knowledge sharing-empowered semantic transmission mechanism in a two-tier edge network, leveraging edge cooperations and bit communications to address KB mismatch. A GESTR was derived and maximized of all MDs by optimizing knowledge sharing decisions, extraction ratios, and BS/subchannel allocations, while meeting task accuracy and delay requirements. The joint optimum solution was obtained for formulated MINLP problem via proposed FP-BnB algorithm and modified K-M algorithm efficiently, which achieved the corresponding performance limit that can be used to shed light on a practical system design. A variety of results help demonstrate the advantages of proposed optimum joint collaborative knowledge sharing design for task-oriented hybrid communications.

\vspace{-2mm}
\bibliographystyle{IEEEtran}
\bibliography{IEEEabrv,mybibfileKBConf}

\end{document}